\documentstyle[pra,aps,multicol,prl,psfig]{revtex}
\renewcommand{\narrowtext}{\begin{multicols}{2} \global\columnwidth20.5pc}
\renewcommand{\widetext}{\end{multicols} \global\columnwidth42.5pc}
\begin{document}
\bibliographystyle{prsty}
\draft

\title{\bf Spin-based quantum information processing with semiconductor quantum dots 
and  cavity QED}
\author{Mang Feng$^{1,2}$, Irene D'Amico$^{1,3}$, Paolo Zanardi$^{1,3}$ and Fausto Rossi$^{1,3,4}$}    
\address{$^{1}$ Institute for Scientific Interchange (ISI) Foundation,\\ 
Villa Gualino, Viale Settimio Severo 65, I-10133, Torino, Italy\\
$^{2}$ Laboratory of Magnetic Resonance and Atomic and Molecular Physics,\\ 
Wuhan Institute of Physics and Mathematics, Chinese Academy of Sciences, Wuhan, 430071, China \\
$^{3}$ Istituto Nazionale per la Fisica della Materia (INFM)\\
$^{4}$ Dipartimento di Fisica, Politecnico di Torino, Corso Duca degli Abruzzi 24-10129 Torino, Italy}
\date{\today}
\maketitle

\begin{abstract}

A quantum information processing  scheme is proposed with semiconductor quantum dots 
located in a high-Q  single mode QED cavity. The spin degrees of freedom of one
 excess conduction electron of the quantum dots are  employed as qubits.
 Excitonic states, which can be produced ultrafastly with optical operation, are 
used as auxiliary states in the realization of quantum gates. 
We show  how    properly tailored  ultrafast laser pulses and Pauli-blocking effects,
can be used to achieve a universal encoded quantum computing.

\end{abstract}
\vskip 0.1cm
\narrowtext

Quantum computing \cite{nielsen} has  drawn much attention over the past few years due to the speedup
it promises  in the treatment of classically hard computational problems, such as factoring 
\cite{shor} and data-base search \cite{grover}. 
Experiments have been made so far in systems of trapped ions, cavity-atom
and nuclear magnetic resonance, which demonstrated the feasibility of small-scale quantum computing
\cite{nielsen}.
However it is generally believed that, in order  to boost   the current techniques to a 
large-scale e.g., thousands of qubits,  quantum computer architecture should be based on solid-state hardware 
exploiting present nanotechnology.

The ideas we will discuss in this paper  are within the framework of semiconductor quantum dot (QD) 
quantum information processing (QIP), which has  been intensively studied by
 envisaging  two different kind of qubit [4-11] based either on  spin or on orbital degrees of freedom. 
In the latter approach
by using the electron-hole pair  states, i.e., excitonic states, as qubits, one 
can have an ultrafast implementation of quantum computing 
with optical operations. The physical coupling between   two (neighboring) qubits  is provided
by dipole-dipole interaction. Decoherence due to phonons is the main obstacle to the implementation
 of this QIP scheme  \cite{biolatti,li}. 
In the former kind of proposals \cite{pazy,loss}, the spin states of the only  excess conduction 
electron of each QD are employed to be qubits. The two-qubit gate is performed on two 
adjacent QDs exploiting   the exchange interaction. This scheme benefits from a much longer 
decoherence time \cite{kikkawa}, but the implementation of quantum gates on spin states is slower
 than that on  excitonic states. A common problem for the two schemes cited above is that 
only the nearest-neighbor  qubits are coupled. So significant overhead is necessary for coupling 
two distant qubits. 
On the other hand, recent developments in semiconductor nanotechnology have shown that 
quantum dots located in high-Q cavity provide an alternative two-level system
 in which the coupling between two distant QDs is mediated by the cavity mode 
\cite{imamo,sherwin}.  So QIP can in principle be implemented  in this kind of systems. 

In the present work, we will try to perform quantum computing with an array of GaAs-based QDs confined in 
a high-Q single mode cavity, by $merging$ the methods of spintronics, 
optoelectronics, and cavity-QED. There is only one excess conduction electron in each QD.
As the cavity mode acts as the 'bus' qubit, two distant qubits can interact directly, which would 
much simplify the quantum computing manipulation. 
Our scheme is inspired by the idea proposed in Ref.\cite{imamo}. In that paper, spin states $m_{x}=1/2$ and 
-1/2 of the only excess conduction electron are employed to be qubit states by applying an additional 
magnetic field along $x$-axis, and an effective long-range interaction is
present between two distant quantum dot spins, mediated by the vacuum field of the cavity mode. 
In our scheme, instead, the magnetic field is applied along $z$-axis. By means of the auxiliary electron-hole
pair states, i.e., excitonic states, we employ the spin states $m_{z}=1/2$ 
and -1/2 of the only conduction electron as qubit states $|1\rangle$ and $|0\rangle$ respectively.
Since excitonic states are introduced as auxiliary states in our scheme, the quantum gates must be
performed quickly because the decoherence time of the exciton is much shorter than that of spin states. 
Moreover, we should also pay attention to the cavity mode, whose decoherence time is of the same order as that
of exciton. Fortunately, as we will show below, both the exciton and the cavity mode are only virtually
excited in our two-qubit gating. Therefore, we can achieve universal quantum computing 
based on a recently proposed model of encoded quantum computing (EQC), in which no single-qubit
operation is needed \cite{lidar}. The experimental feasibility of our scheme will also be discussed.  

We assume that, besides radiated by the cavity light, the QDs can be individually addressed by lasers. 
Due to Pauli exclusion principle, the radiation of a $\sigma^{-}$ polarized light with suitable energy on 
the QD with projected angular momentum of $\pm\frac {1}{2}$
(in the unit of $\hbar=1$) will produce an exciton with state  
$|m_{J}^{e}=-\frac {1}{2}, m_{J}^{h}=-\frac {1}{2}\rangle$ in the s-shell only if the excess electron has
a spin projection $\frac {1}{2}$. This Pauli-blocking mechanism has been observed experimentally in QDs 
\cite{bonadeo,chen} and can be used to produce entangled states. In Ref.\cite{pazy}, this Pauli-blocking was
used to yield a conditional phase gate, together with the Coulomb interaction between two neighboring QDs. 
In the single-particle picture, we define 
$|0\rangle_{\nu}=c^{\dagger}_{\nu,0,-\frac {1}{2}}|vac\rangle$, 
$|1\rangle_{\nu}=c^{\dagger}_{\nu,0,\frac {1}{2}}|vac\rangle$, and the excitonic state
$|X^{-}\rangle_{\nu}=c^{\dagger}_{\nu,0,-\frac {1}{2}}c^{\dagger}_{\nu,0,\frac {1}{2}}
d^{\dagger}_{\nu,0,-\frac {1}{2}}|vac\rangle$, where $c^{\dagger}_{\nu,i,\sigma} (d^{\dagger}_{\nu,i,\sigma})$
is the creation operator for a conduction (valence) band electron (hole) in the $i$-th single particle state 
of QD $\nu$, with spin projection $\sigma$, and $|vac\rangle$ accounts for the excitonic vacuum. 
The Hamiltonian of the QDs system is generally written as
\begin{equation}
H=\hbar\omega_{c}a^{\dagger}a+\sum_{k} H_{k}+\sum_{k} H^{int}_{k}
\end{equation}
where $\omega_{c}$ is the cavity frequency, $a^{\dagger}$ and $a$ are creation and annihilation
operators of the cavity. $H_{k}$ is the single-QD Hamiltonian composed of $H^{0}_{k}$ and $H^{co}_{k}$, with 
$H^{0}_{k}=\sum_{i,\sigma=\pm 1/2} \epsilon^{e}_{i\sigma} c^{\dagger}_{k i
\sigma}c_{k i \sigma} + \sum_{j,\sigma'=\pm 1/2} \epsilon^{h}_{j\sigma'}
d^{\dagger}_{k j \sigma'}d_{k j \sigma'}$
describing the  independent electrons and holes in the QDs, in which
$\epsilon^{e}_{i\sigma}$ and  $\epsilon^{h}_{j\sigma'}$ are respectively
eigenenergies of an electron with spin projection $\sigma$ in the  $i$-th
single particle state of QD $k$ and a hole with spin projection $\sigma'$ in
the j-th single particle  state of QD $k$. $H^{co}_{k}$ is the electron-hole
Coulomb interaction. $H^{int}_{k}=H_{k}^{L}+H_{k}^{c}$ with
$H^{L}_{k}$ and $H^{c}_{k}$ being the laser-QD interaction and  cavity-QD
interaction respectively.  

Two-qubit gate performance is the focus of various quantum computing proposals. As QDs are put into the
cavity, the two spin states, employed as qubits, can be coupled via the cavity mode. Let us first  
consider the QD $k$, which is radiated by cavity light with $\sigma^{+}$ polarization and laser beam 
with linear polarization, as shown in Fig.1 where the energy difference between the conduction band electron
and the valence band hole in the excitonic state $|X^{-}\rangle$ is $\hbar\omega_{0}^{k}$, the cavity
frequency $\omega_{c}=\omega_{d}^{k}+\omega_{0}^{k}-\Delta_{k}-\delta_{k}$ and laser frequency
$\omega_{L}^{k}=\omega_{d}^{k}-\Delta_{k}$. Both $\delta_{k}$ and $\Delta_{k}$ are detunings, where
$\delta_{k}$ can be written as $\omega_{L}^{k}+\omega_{0}^{k}-\omega_{c}$. If 
$\delta_{k}\rightarrow 0$ and $\Delta_{k}$ is large enough, then we have a typical resonance Raman
transition between $|1\rangle$ and $|0\rangle$, whose interaction Hamiltonian in the unit of $\hbar=1$ is
\begin{equation}
H_{int}= \frac {\Omega_{k}(t)}{2} [a \sigma_{01}^{k}e^{i\omega_{L}^{k}t} + h.c.]
\end{equation}
with $\Omega_{k}(t)=G_{c}G^{k}_{las}(t)[1/\Delta_{k}+1/(\Delta_{k}+\delta_{k})]$, $G_{c}$ and 
$G^{k}_{las}(t)$ being cavity-QD  and laser-QD couplings, respectively. $\sigma_{01}^{k}=|1>_{k}<0|$ and  
no excitation in state $|-1/2\rangle_{h}$. From now on, we consider two identical 
QDs A and B, and set $\omega_{d}^{A}=\omega_{d}^{B}$ and $\omega_{0}^{A}=\omega_{0}^{B}=\omega_{0}$.
If we set $\omega_{L}^{A}=\omega_{L}^{B}$, then we have $\Delta_{A}=\Delta_{B}$ and 
$\delta_{A}=\delta_{B}=\delta$. To suppress the cavity decay as much as we can, in the remainder of the paper,
we suppose that the cavity mode is in vacuum state.
By adjusting cavity light and laser beam to make $\delta$ smaller than 
$\omega_{0}$, but larger than both $\Omega_{k}(t)$ and cavity linewidth, we will have a near two-photon 
resonance condition for two qubits, with the following effective Hamiltonian under the rotating-wave 
approximation \cite{butcher},
\begin{equation}
H_{eff}=\frac {\tilde{\Omega}(t)}{2} (\sigma_{01}^{A}\sigma_{01}^{\dagger B} + 
\sigma_{01}^{B}\sigma_{01}^{\dagger A}),
\end{equation}
where $\tilde{\Omega}(t)= \Omega_{A}(t)\Omega_{B}(t)/(2\delta)$.  
By means of Eq.(3), we may obtain the time evolution of the system,
\begin{equation}
|01\rangle_{AB} \rightarrow \cos [\frac {1}{2}\int_{0}^{T} \tilde{\Omega}(t) dt] |01\rangle_{AB} - i 
\sin [\frac {1}{2}\int_{0}^{T} \tilde{\Omega}(t) dt] |10\rangle_{AB},
\end{equation}
and 
\begin{equation}
|10\rangle_{AB} \rightarrow \cos [\frac {1}{2}\int_{0}^{T} \tilde{\Omega}(t) dt] |10\rangle_{AB} - i 
\sin [\frac {1}{2}\int_{0}^{T} \tilde{\Omega}(t) dt] |01\rangle_{AB}
\end{equation}
with $|\cdot\cdot\rangle_{AB}$ being the product of internal states of QDs A and B. It means that, no matter 
whether QDs A and B are adjacent or not, their internal states can be entangled by
coupling to the same cavity mode, although the cavity mode is only virtually populated.
Eq.(3) is also called XY model. Based on it, a universal EQC can be constructed by means of
the nearest-neighbor and next-nearest-neighbor couplings \cite{lidar}.  
The idea is to encode logical qubits in the state-space of pairs of adjacent QDs:
$|0_L>_i:= |01\rangle_{i,i+1},\,|1_L>_i:= |10\rangle_{i,i+1}.$
Given this encoding Wu and Lidar showed in Ref. \cite{lidar} how arbitrary
qubits manipulations i.e., universality, can be achieved just by time-dependent control
of the XY hamiltonian with nearest-neighbor and next-nearest-neighbor interactions.
The necessity of the difficult single-qubits operation is relaxed by this way.
This scheme fits in the general conceptual framework of {\em encoded universality} (see again
\cite{lidar} and references therein) in which one exploits the naturally available interactions 
in the system, in such a way to enact universality in a suitable subspace i.e., the code,
of the full physical state-space.
Notice that  our scheme meets the
requirement of EQC if Coulomb interaction can be
neglected due to large enough distance between two neighboring QDs. When EQC is performed in our
scheme, however, the short decoherence time of the excitonic state must be
seriously considered. Besides,
the cavity decay has also a detrimental effect on our scheme although cavity mode is factorized from the
computational subspace. This is because the fluctuation of the cavity mode would affect the 'bus' role it 
plays and therefore affects the coupling of the two distant spin qubits.    
Consequently the implementation time of Eq.(3) is required to be shorter than the
decoherence time of the cavity mode and the excitonic state. In order for Eq.(3) to work, the following 
adiabatic conditions must be fulfilled:
\begin{eqnarray}
\Delta_{k}&\gg & \delta_{k}  \gg  max({\Omega_{k}\over 2},{1\over\tau})\label{first}\\
\Delta_{k}+\delta_{k} & \gg & max( G_c,{1\over\tau}) \\
\Delta_{k} & \gg & max({ G^k_{las}},{1\over\tau}),\label{last}  
\end{eqnarray}
where $\tau$ is the characteristic time associated to a Gaussian laser pulse of the form  
$G^k_{las}(t)=G^k_{las}\exp(-t^2/ 2\tau^2)$. By analyzing the whole parameter space while imposing (i)
conditions (\ref{first})-(\ref{last}) and (ii) $\int_{0}^{T} \tilde{\Omega} (t)dt =2\pi$, we obtain that the 
points available to our computation in the parameter plane $(G_c,\tau)$ are the ones corresponding to the 
shaded region in Fig. 2. In particular if we consider a coupling strength $G_{c}$ of the order of 1 meV, we 
see that the characteristic time associated to the implementation of Eq. (3) will be of the order of 150 ps. 
Fortunately, in the implementation of Eq. (3), both the cavity mode and the exciton are only virtually 
excited. If we suppose 
that the probability of their excitations is less than 1$\%$ \cite{imamo}, the coherent implementation time 
of Eq.(3) can be at least 100 times longer than the decoherence time of the
cavity and the exciton themselves, i.e., as long as 1 $ns$. This implies that Eq. (3) will work well.

We will now compare our scheme with previous ones involving spin qubits.
The obvious difference of our scheme from Ref.\cite{pazy} is that the two QDs are interacted via the cavity 
mode, instead of the Coulomb interaction. So the biexcitonic shift produced in Ref.\cite{pazy} by the 
Coulomb interaction between two QDs is not necessary any more and the external in-plane electric field 
applied to enlarge the biexcitonic shift can be removed. Moreover, the two-qubit gate implemented on two 
non-neighboring QDs makes our scheme of quantum computing more efficient than those proposals based on the 
nearest-neighbor coupling \cite{pazy,loss}. It is also the prerequisite of our scheme applicable to EQC.
Furthermore, our scheme is different from Ref.\cite{imamo}. As the Pauli blocking is introduced, we 
employ the spin states of $m_{z}=\pm 1/2$ to be qubits. Due to this fact, we can perform Eq.(3) without any 
external magnetic field \cite{exp1}. 

For achieving the scheme experimentally, III-V semiconductor material is a suitable candidate 
because of the low spin decoherence rate of conduction electron. Each QD must be 
initially cooled and prepared to contain only one excess electron. As far as we know,
this has been experimentally achieved \cite{fin}.  
Moreover individual addressing of QD by laser beam is necessary, which is a challenge for almost all 
proposals of semiconductor quantum computing. But in our scheme, as Coulomb interaction is not necessary, a 
possible way to avoid this difficulty is to enlarge the spacing between two adjacent QDs and use near field 
techniques. Furthermore, to perform quantum computing in parallel in cavity-QED, it is generally required that 
the decoherence time of the cavity photon must be very long. However, this requirement can be removed because
the cavity mode is only virtually populated throughout our scheme.
For the measurement of the final result, we can adopt the method
proposed in \cite{imamo} by employing the Raman transition between $|1\rangle$ and $|0\rangle$. If the QD 
spin state is in $|1\rangle$, a photon would be created in the cavity and eventually leak out of the cavity. 
So by detecting the single photon signal, we can judge whether the QD spin state is in $|1\rangle$ or 
$|0\rangle$.

The quantum gate based on our scheme can be carried out with high fidelity. To our knowledge, possible
sources of error are ($i$) there is probably a small admixture of heave
hole component to the light hole wavefunction, which yields the excitonic state 
$|m^{e}_{J}=-1/2, m^{h}_{J}=3/2\rangle$ in each cavity radiation with the $\sigma^{+}$ polarization 
in the case that the spin projection of the only excess electron is $+\frac {1}{2}$.
 To avoid this situation, we can adjust the strength of the magnetic field to make the 
radiated light non-resonant with the undesired transition.  So it is
expected that the probability of this error would be very small; ($ii$) when EQC is performed, 
the F{\"o}rster process \cite{quiroga} happening in the nearest-neighbor coupled QDs would probably take 
place. But due to both the spin-selection rule and energy-conservation rule, and in particular, the 
relatively large distance
between two neighboring QDs, this kind of process would be largely inhibited. 

In summary, we have reported an EQC scheme of quantum computing with semiconductor QDs in a high-Q single mode
cavity. The experimental feasibility of implementing our scheme has been discussed based on our numerical
estimate for the adiabatic manipulation of two-qubit gate. To minimize the gating 
time, a stronger coupling between the dots and the cavity is expected. In principle, our scheme can 
be generalized to the many-qubit case, in which quantum gates are performed in parallel. 
However, we should note that to implement EQC, we need twice of the qubits and more operations compared to 
non-encoded quantum computing schemes, 
which is also a challenge for current cavity QED experiment. For example, with current cavity technique, the 
larger the cavity size, the weaker the coupling strength and the shorter the decoherence 
time of the cavity system. Actually, Eq. (3) is also very useful in usual quantum computing schemes, in which
single-qubit operation is needed. For the system under consideration, we may easily perform the 
single-qubit rotation by two lasers with different polarizations and suitable frequencies \cite{imamo} to 
meet the Raman-resonance condition between $|1\rangle$ and $|0\rangle$. Alternatively, we may rotate the spins
by laser pulses, assisted by a magnetic field, as proposed recently in an ultrafast manipulation method
\cite{gupta}. This means that our approach resulting in Eq. (3) is applicable to various non-encoded quantum
computing schemes. 

The work is supported by the European Commission through the Research Project SQID within FET Program.

\begin{figure}
 \label{16thnewfig1}
 \psfig{figure=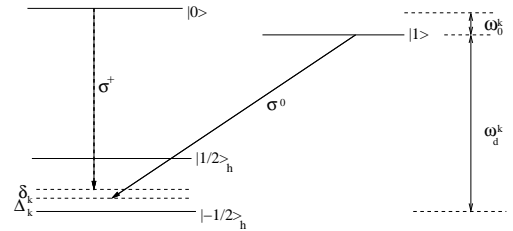,width=1.20\columnwidth,angle=-90}
  \caption{
  Configuration of the quantum dot $k$ in the near two-photon resonance process, where 
$|0\rangle=|-1/2\rangle_{e}$, $|1\rangle=|1/2\rangle_{e}$.
$\omega_{c}$ and $\omega_{L}^{k}$ are frequencies of the cavity and the laser respectively. 
The cavity light is $\sigma^{+}$ polarized and the laser beam is of linear polarization. $\Delta_{k}$ and 
$\delta_{k}$ are detunings defined in the text. }
\end{figure}

\begin{figure}
  \label{tau_ca_det}
\psfig{figure=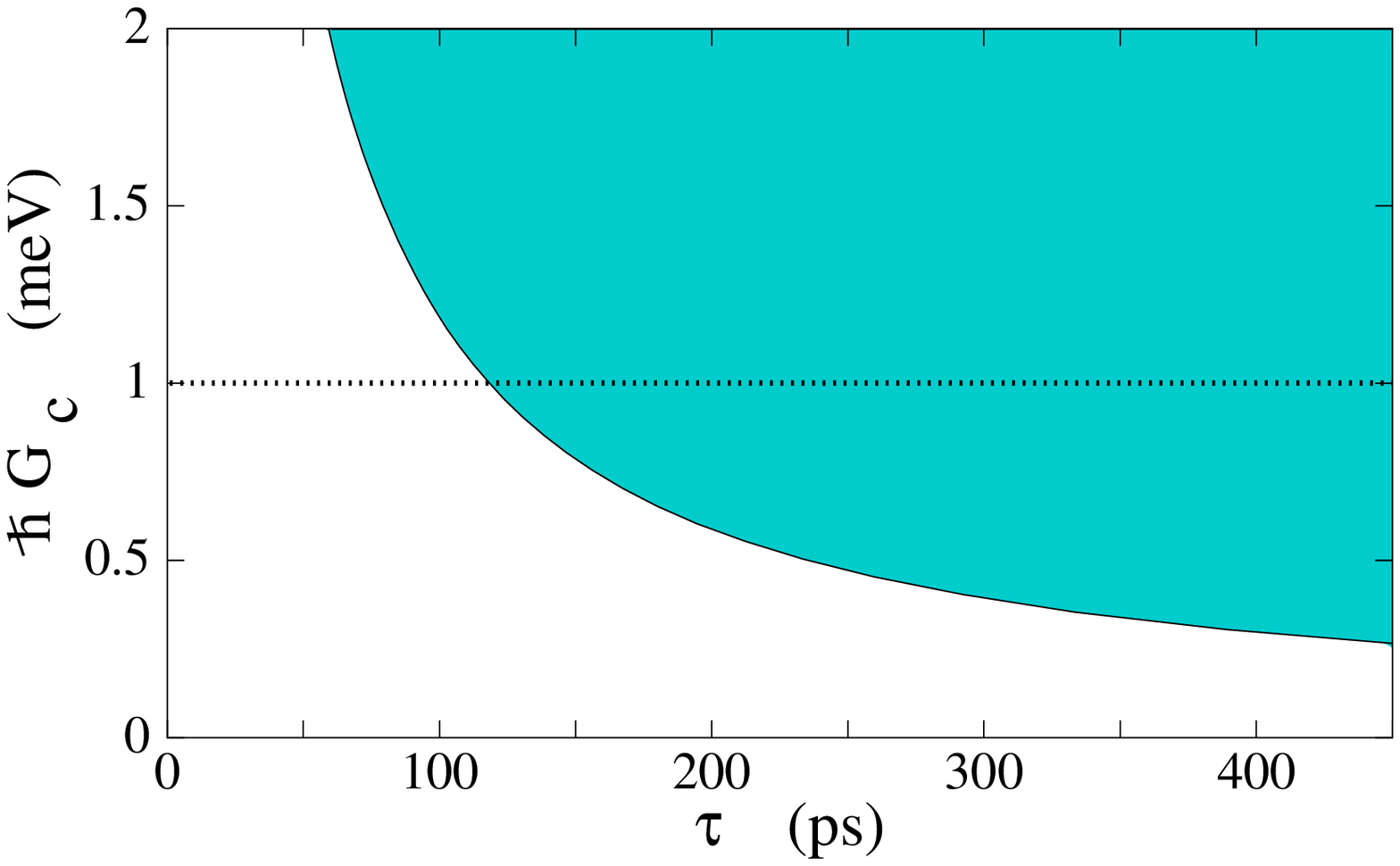,width=1.20\columnwidth,angle=-90}
  \caption{Plot of the parameter space available (shaded area) for Eq. (3) in the 
  implementation of $\int_{0}^{T} \tilde{\Omega} (t)dt =2\pi$.}
\end{figure}

\widetext

\begin{thebibliography}{24}
\bibitem{exp1} In the text above and in Fig.1, the effect of the external magnetic field is taken into 
account, because the magnetic field along $z$-axis is needed to remove the degeneracy of the qubit states 
when EQC is carried out. But if we only consider the implementation of Eq.(3) or any other non-encoded 
quantum computing scheme, such a magnetic field may be unnecessary.
\end{thebibliography}
\end{document}